# Applications of Lambert-Tsallis and Lambert-Kaniadakis Functions in Differential and Difference Equations with Deformed Exponential Decay

J. L. E. da Silva     G. B. da Silva     R. V. Ramos

leonardojade@alu.ufc.br     george_barbosa@fisica.ufc.br     rubens.ramos@ufc.br

*Lab. of Quantum Information Technology, Department of Teleinformatic Engineering – Federal University of Ceara - DETI/UFC, C.P. 6007 – Campus do Pici - 60455-970 Fortaleza-Ce, Brazil.*

*Abstract*

The analysis of a dynamical system modelled by differential (continuum case) or difference equation (discrete case) with deformed exponential decay, here we consider Tsallis and Kaniadakis exponentials, may require the use of the recently proposed deformed Lambert functions: the Lambert-Tsallis and Lambert-Kaniadakis functions. In this direction, the present work studies the logistic map with deformed exponential decay, using the Lambert-Tsallis and the Lambert-Kaniadakis functions to determine the stable behaviour and the dynamic of the disentropy in the weak chaotic regime. Furthermore, we investigate the motion of projectile when the vertical motion is governed by a non-linear differential equation with Tsallis exponential in the coefficient of the second order derivative. In this case, we calculated the range of the projectile using the Lambert-Tsallis function.

## 1. Introduction

The Lambert *W* function is an important elementary mathematical function that finds applications in different areas of mathematics, computer Science and physics [1-6]. Basically, the Lambert *W* function is defined as the solution of the equation

$$W(z)e^{W(z)} = z. \tag{1}$$

Since *W*(*z*) is a non-injective function, there exist infinite solutions, however, only two of them returns a real value when the argument *z* is real. In the interval $-1/e \leq x \leq 0$ there exist two real values of *W*(*z*). The branch for which $W(x) \geq -1$ is the principal branch named $W_0(z)$ while the branch satisfying $W(z) \leq -1$ is named $W_{-1}(z)$. For $x \geq 0$ only $W_0(z)$ is real and for $x < -1/e$ there are not real solutions. The point ($z_b = -1/e$, $W(z_b) = -1$) is the branch point where the solutions $W_0$ and $W_{-1}$ have the same value and $dW/dz = \infty$. The plot of *W*(*z*) versus *z* is shown in Fig. 1.

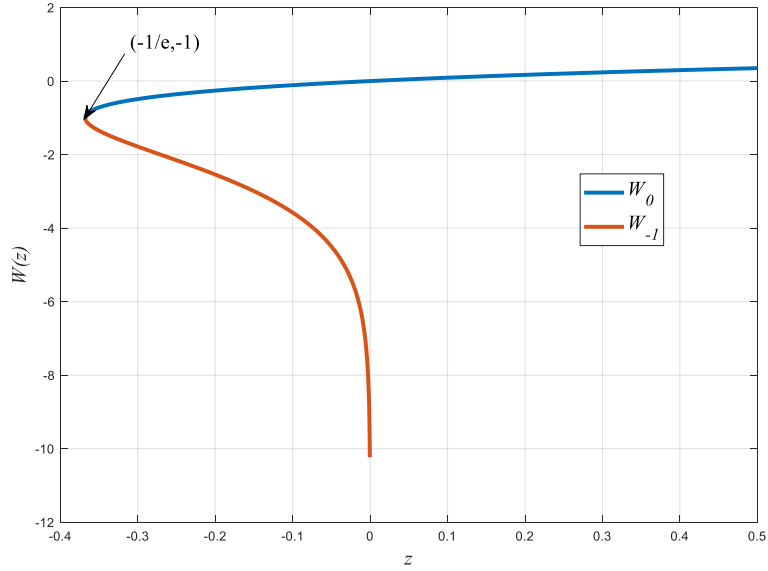

Fig. 1 – $W(z)$ versus $z$

It can be easily shown that

$$W(z_1)+W(z_2) = W\left(z_1 z_2 \left(\frac{W(z_2)+W(z_1)}{W(z_1)W(z_2)}\right)\right) \qquad (2)$$

$$\frac{dW}{dz} = \frac{W(z)}{z[W(z)+1]}, \quad z \neq 0. \qquad (3)$$

The most recent application of the Lambert $W$ function is in the calculation of the disentropy [7]:

$$D = \sum_i p_i W(p_i), \qquad (4)$$

where $\{p_1, p_2, \ldots, p_n\}$ is a probability distribution. When the disentropy is minimal entropy is maximal and vice-versa. Equation (4) is the disentropy related to Boltzmann-Gibbs entropy. The disentropy related to Shannon entropy is

$$D = \sum_i p_i R_2(p_i) \qquad (5)$$

$$R_2(z) = \log_2(e) W\left(\frac{z}{\log_2(e)}\right). \tag{6}$$

Aiming to model natural systems that do not follow a pure exponential/logarithmic law, deformations of the exponential and logarithmic functions have been proposed. This allowed the deformation of the Gaussian distribution and the generalization of the von Neumann entropy, for example. The most famous deformations with large applications in physics appeared with the proposal of Tsallis $q$-entropy [8] and Kaniadakis $\kappa$-entropy [9]. The $q$-exponential and $q$-logarithm functions are given by

$$e_q^z = \begin{cases} e^z & q = 1 \\ [1+(1-q)z]^{1/(1-q)} & q \neq 1 \ \& \ 1+(1-q)z \geq 0 \\ 0 & q \neq 1 \ \& \ 1+(1-q)z < 0 \end{cases} \tag{7}$$

$$\ln_q(z) = \begin{cases} \ln(z) & x > 0 \ \& \ q = 1 \\ \dfrac{x^{(1-q)}-1}{1-q} & x > 0 \ \& \ q \neq 1 \\ \text{undefined} & x \leq 0 \end{cases}. \tag{8}$$

Furthermore $\exp_q(\ln_q(z)) = z$ for $z > 0$ and $\ln_q(\exp_q(z)) = z$ for $0 < \exp_q(z) < \infty$.

On the other hand the $\kappa$-exponential and $\kappa$-logarithmic functions are given, respectively, by

$$e_\kappa^z = \left[\sqrt{1+\kappa^2 z^2} + \kappa z\right]^{\frac{1}{\kappa}} \tag{9}$$

$$\ln_\kappa(x) = \frac{x^\kappa - x^{-\kappa}}{2\kappa}. \tag{10}$$

Those deformations give rise to $q$-algebra [10] and $\kappa$-algebra [9,11].

After the deformation of exponential and logarithmic functions it is natural to consider the deformation of the Lambert $W$ function. In fact, two deformations were recently proposed, the Lambert-Tsallis $W_q$ function [7] and the Lambert-Kaniadakis $W_\kappa$

function [12]. They are, respectively, the solutions of

$$W_q(z)e_q^{W_q(z)} = z \qquad (11)$$

$$W_\kappa(z)e_\kappa^{W_\kappa(z)} = z. \qquad (12)$$

Using the definition of $exp_q$ given in eq. (7), the function $W_q$ can be found solving the equation [7]

$$x(r+x)_+^r = r^r z, \qquad (13)$$

where $x = W_{\frac{(r-1)}{r}}(z)$, $\frac{1}{1-q} = r$ and $(a)_+ = \max\{a, 0\}$. When $q = 1$, one has $e_q(z) = e^z$ and, consequently, $W_1(z) = W(z)$.

It can be shown the branch point of the Lambert-Tsallis $W_q$ function is ($z_b = exp_q(1/(q\text{-}2))/(q\text{-}2)$, $W_q(z_b) = 1/(q\text{-}2)$), for $q \neq 2$. There is no branch point with finite $z_b$ for $q = 2$. For $q = 1$, the branch point of Lambert $W$ function is recovered. The solution in the interval $z_b \leq z < 0$ is $W_q^-(z)$ while the solution in the interval $z_b \leq z < \infty$ is $W_q^+(z)$. The function $W_q^+(z)$ keeps its concavity according to $d^2 W_q^+(z)/dz^2 < 0$. On the other hand, $W_q^-(z)$ decreases from the branch point and goes toward the point $(0^-, -\infty)$. For example, for $q = \{2, 3/2, 1/2\}$ one has the following upper branches

$$W_2(z) = \frac{z}{z+1}, \qquad z > -1, \qquad (14.\text{a})$$

$$W_{3/2}^+(z) = \frac{2(z+1)+2\sqrt{2z+1}}{z}, \qquad z > -1/2, \qquad (14.\text{b})$$

$$W_{1/2}^+(z) = \frac{\left[3\sqrt[3]{2z+\sqrt{\left(2z+\frac{8}{27}\right)^2 - \frac{64}{729}} + \frac{8}{27}} - 2\right]^2}{9\sqrt[3]{2z+\sqrt{\left(2z+\frac{8}{27}\right)^2 - \frac{64}{729}} + \frac{8}{27}}}, \qquad z \geq -0.29629, \qquad (14.\text{c})$$

Figure 2 shows the plot of $W_{q=3/2}$ versus $z$.

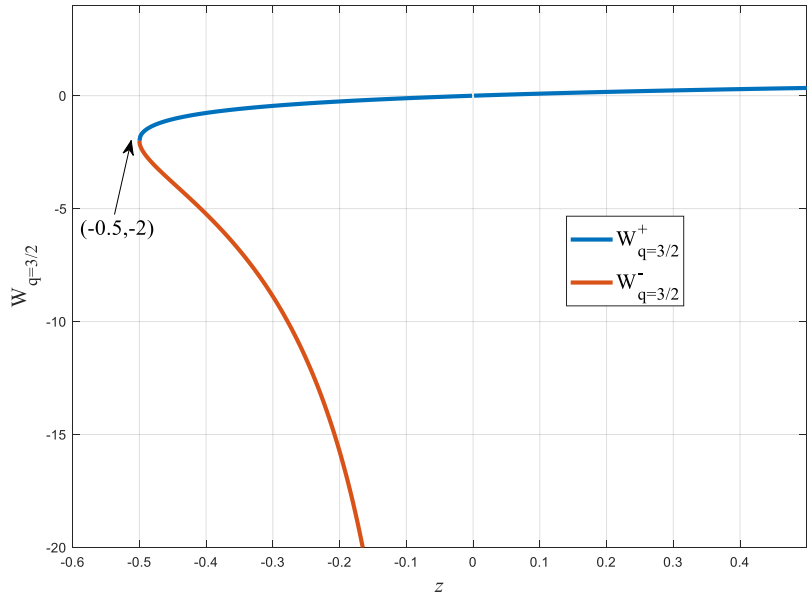

Fig.2. $W_{q=3/2}$ versus $z$.

Similarly, using the definition of $exp_\kappa$ given in eq. (9), $W_\kappa$ can be found solving the equation [12]

$$x\left[\sqrt{r^2+x^2}+x\right]^r = r^r z, \qquad (15)$$

where $r = 1/\kappa$ and $x = W_{\frac{1}{r}}(z)$. Obviously $W_{\kappa=0}(z) = W(z)$. The branch point is ($z_b$ = -(1-$\kappa$)$^{(1-\kappa)/2\kappa}$/(1+$\kappa$)$^{(1+\kappa)/2\kappa}$, $W_\kappa(z_b)$ = -(1-$\kappa^2$)$^{-1/2}$), that is valid in the interval $0 \leq \kappa^2 < 1$. Thus the solution in the interval $z_b \leq z < 0$ is $W_\kappa^-(z)$ while the solution in the interval $z_b \leq z < \infty$ is $W_\kappa^+(z)$. For example, for $\kappa = 1/3$ eq. (15) assumes the form

$$\left(x^{2/3}\right)^2 + \frac{3}{2z^{1/3}}x^{2/3} - \frac{3}{2}z^{1/3} = 0. \qquad (16)$$

Hence, the Lambert-Kaniadakis functions for $\kappa = 1/3$ are

$$W_{1/3}^{\pm}(z)=\left(-\frac{3}{4z^{1/3}}\pm\frac{3}{4z^{1/3}}\sqrt{1+\frac{8}{3}z}\right)^{3/2}. \tag{17}$$

On the other hand, for $\kappa = 1/5$ the Lambert-Kaniadakis functions are two of the three roots of

$$\left(x^{2/5}\right)^3+\frac{5}{2z^{1/5}}x^{2/5}-\frac{5}{2}z^{1/5}=0. \tag{18}$$

The plot of $W_{\kappa=1/5}$ can be seen in Fig. 3.

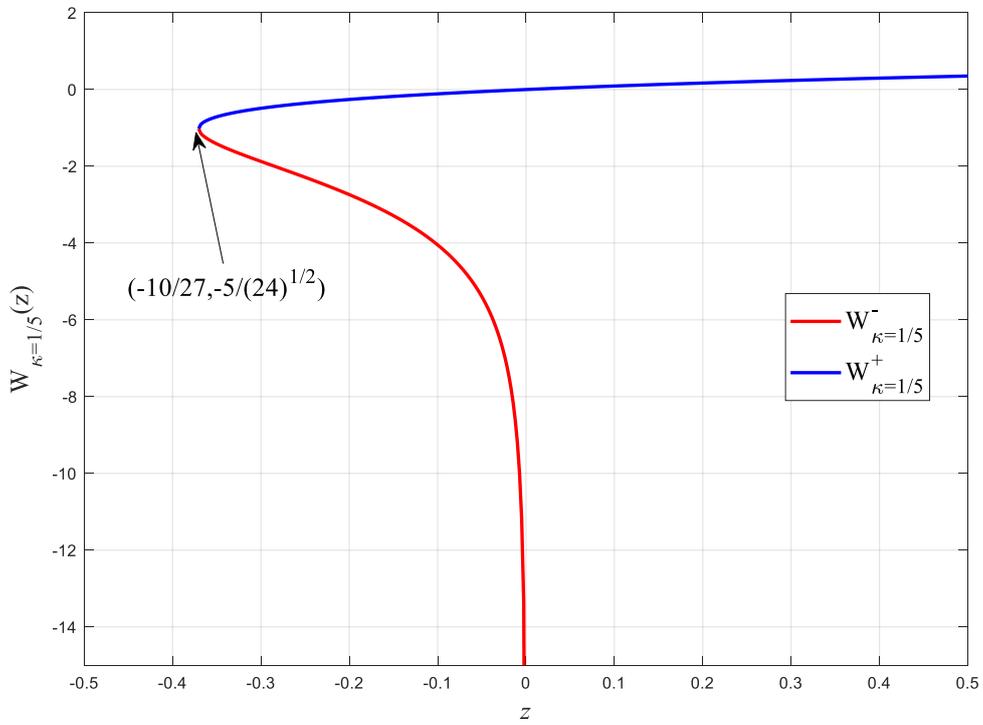

Fig. 3. $W_{\kappa=1/5}$ versus $z$.

Similar to eq. (2) one can show that

$$W_q(z_1)\oplus_q W_q(z_2)=W_q\left[z_1z_2\left(\frac{W_q(z_1)\oplus_q W_q(z_2)}{W_q(z_1)W_q(z_2)}\right)\right] \tag{19}$$

$$W_\kappa(z_1) \oplus_\kappa W_\kappa(z_2) = W_\kappa\left[z_1 z_2 \left(\frac{W_\kappa(z_1) \oplus_\kappa W_\kappa(z_2)}{W_\kappa(z_1)W_\kappa(z_2)}\right)\right]. \tag{20}$$

An important property of the $W$ function that extends to $W_{q(\kappa)}$ function is the convergence of the limit $\lim_{z \to 0} W(z)/z = 1$, for $z$ real. In fact, from eq. (13) and eq. (15) one has $W_{q(\kappa)}(0) = 0$, hence $\lim_{z \to 0} \exp_{q(\kappa)}[W_{q(\kappa)}(z)] = 1$. Using the last in eqs. (11) and (12), one has

$$\lim_{z \to 0} \frac{W_{q(\kappa)}(z)}{z} \exp_{q(\kappa)}\left[W_{q(\kappa)}(z)\right] = 1 \Rightarrow \lim_{z \to 0} \frac{W_{q(\kappa)}(z)}{z} = 1. \tag{21}$$

Furthermore, using their definitions, one also has

$$\lim_{z \to +\infty} \frac{W_{q(\kappa)}(z)}{z} = 0. \tag{22}$$

After some algebra, one can show the first derivatives of $W_q$ and $W_\kappa$ are given by

$$\frac{dW_q(z)}{dz} = \frac{-\left((1-q)W_q(z)+1\right)^{\frac{q}{q-1}}}{(q-2)W_q(z)-1} \tag{23}$$

$$\frac{dW_\kappa}{dz} = \frac{\sqrt{1+\kappa^2 W_\kappa^2}}{\exp_\kappa(W_\kappa)\left(\sqrt{1+\kappa^2 W_\kappa^2}+W_\kappa\right)} \tag{24}$$

The coordinates of the branches points of the functions $W_{q(\kappa)}(z)$ are obtained doing $dW_{q(\kappa)}/dz = \infty$ (points of the curve with vertical tangent). The first solution is $W_{q(\kappa)} = -\infty$. The second solution, $W_{q(\kappa)} = W_{q(\kappa)}^b$, depends on the value of $q(\kappa)$, as shown before.

## 2. Difference equation with $q(\kappa)$-exponential decay

Let us initially consider the $q(\kappa)$-version of the nonlinear map used in the study of red blood cells survival described in [13]

$$x_{n+1} = (1-\alpha)x_n + \frac{p}{e_{q(\kappa)}^{\sigma x_n}}, \qquad (25)$$

where $p, \sigma > 0$ and $\alpha \in (0,1)$. The stable solution requires $x_{n+1} = x_n = x$. Hence, the stable solution can be found as follows,

$$x = (1-\alpha)x + \frac{p}{e_{q(\kappa)}^{\sigma x}} \Rightarrow \sigma x e_{q(\kappa)}^{\sigma x} = \frac{\sigma p}{\alpha} \Rightarrow x = \frac{W_{q(\kappa)}(\sigma p/\alpha)}{\sigma}. \qquad (26)$$

For example, setting $p = 1$, $\alpha = 0.9$ and $q = 2$ in eq. (25) one can observe in Fig. 4 the stable regime for $\sigma = 0.5$ (top) and the periodic regime for $\sigma = 1.1$ (bottom). In the stable regime the map converges numerically to the value $x = 0.529105$ while the value provided by eq. (26) is $x = 0.529104$.

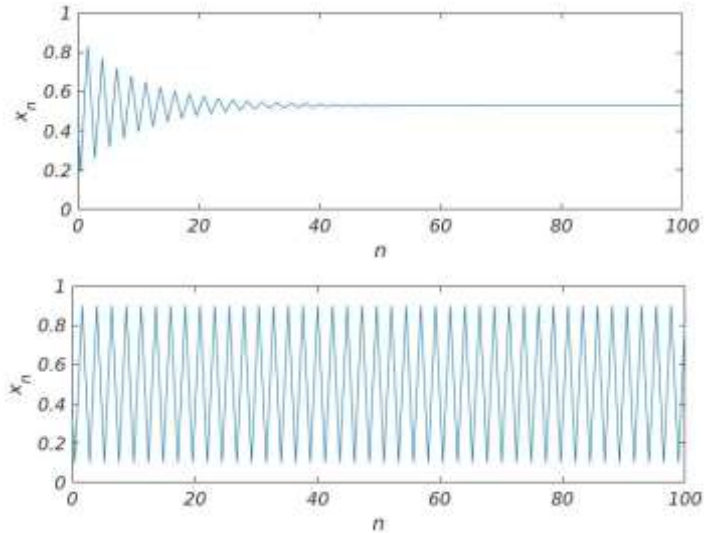

Fig. 4. Iterations of eq. (25) for $p = 1$, $\alpha = 0.9$ and $q = 2$: $\sigma = 0.5$ (stable solution), $\sigma = 1.1$ (periodic solution).

On the other hand, using $\kappa = \frac{1}{2}$, $p = 50$ and $\sigma = 20$, one gets a stable solution for $\alpha = 0.2$ and a periodic solution for $\alpha = 0.85$, as shown in Fig. 5.

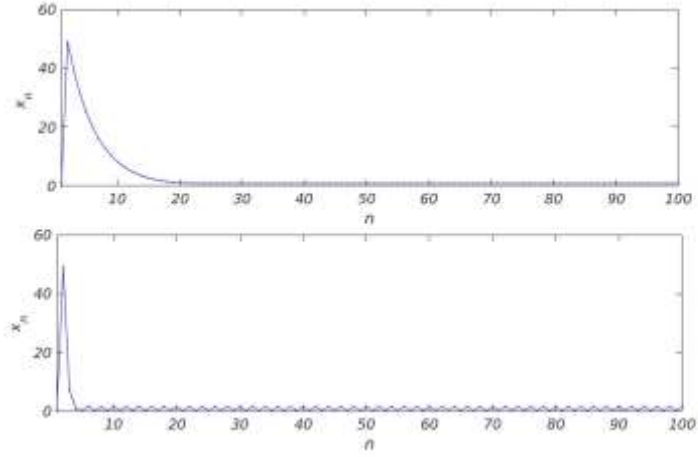

Fig. 5. Iterations of eq. (25) for $p = 50$, $\sigma = 20$ and $\kappa = 1/2$: $\alpha = 0.2$ (stable solution), $\alpha = 0.85$ (periodic solution).

For the stable case, the value obtained numerically and the value provided by eq. (26) are the same: $x = 0.85304$.

Now, let us consider the logistic map with a deformed exponential decay,

$$x_{n+1} = \lambda x_n (1 - x_n) + \frac{P}{e_{q(\kappa)}^{\lambda x_n^2 + (1-\lambda)x_n}}. \tag{27}$$

Substituting $x_{n+1} = x_n = x$ in eq. (27) one gets

$$\{\lambda x^2 + (1-\lambda)x\}.e_{q(\kappa)}^{\lambda x^2 + (1-\lambda)x} = P \Rightarrow x^2 + \frac{(1-\lambda)}{\lambda}x - \frac{W_{q(\kappa)}(P)}{\lambda} = 0 \tag{28}$$

Hence, if eq. (27) reaches a stable solution, its value will be one of the roots of eq. (28):

$$x = \frac{1}{2}\left[-\left(\frac{\lambda-1}{\lambda}\right) \pm \sqrt{\left(\frac{1-\lambda}{\lambda}\right)^2 + \frac{4W_{q(\kappa)}(P)}{\lambda}}\right]. \tag{29}$$

For example, for $\kappa = 0.75$ and $\lambda = 0.5$, the map given in eq. (27) converges to the positive root of (29): $x = 0.68158$. Similarly, for $q = 1.5$ and $\lambda = 0.2$, the map given in eq. (27) converges to the positive root of (29): $x = 0.58447$.

At last, let us consider the following map.

$$x_{n+1} = \lambda x_n(1-x_n) + \frac{P}{e_{q(\kappa)}^{\sigma x_n}}. \tag{30}$$

Differently of the maps given in eqs. (25) and (27), the stable solution cannot be directly obtained by just doing $x_{n+1} = x_n = x$ in eq. (30). However, using the $W_{q(\kappa)}$ function in eq. (30) one can easily get the following map

$$x_{n+1} = \frac{1}{\sigma} W_{q(k)}\left(\frac{\sigma P}{[(1-\lambda)+\lambda x_n]}\right). \tag{31}$$

The maps in eqs. (30) and (31) have the same stable solution (although they do not have the same dynamic). This can be seen in the bifurcation diagram ($p = 1$, $\sigma = 0.9$, $q = 2$) shown in Fig. 6. In the interval $0 \leq \lambda \leq {\sim}0.53$ the maps in eqs. (30) and (31) converges to the same value.

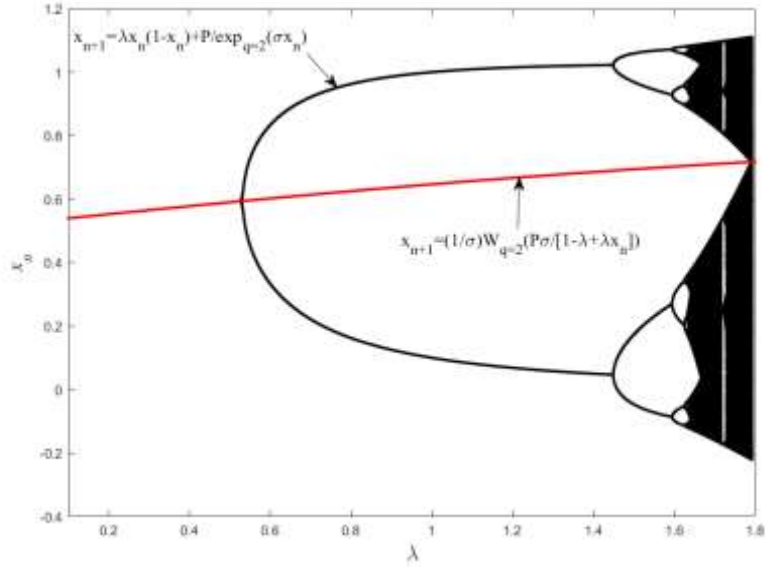

Fig. 6. Bifurcation diagram of the maps given in eqs. (30) and (31) for $p = 1$, $\sigma = 0.9$ and $q = 2$.

The Lyapunov numbers of the logistic map with deformed exponential decay given in eq. (30) are

$$\Lambda(\lambda) = \lim_{N \to \infty} \frac{1}{N} \sum_{n=0}^{N} \ln\left|\lambda - 2\lambda x_n - \frac{P\sigma(1+(1-q)\sigma x_n)^{\frac{1}{1-q}-1}}{\left(e_q^{\sigma x_n}\right)^2}\right|, \tag{32}$$

$$\Lambda(\lambda) = \lim_{N \to \infty} \frac{1}{N} \sum_{n=0}^{N} \ln \left| \lambda - 2\lambda x_n - \frac{P\sigma e_\kappa^{\sigma x_n}}{\sqrt{1 + (\kappa \sigma x_n)^2 \left(e_\kappa^{\sigma x_n}\right)^2}} \right|. \qquad (33)$$

The bifurcation diagram and Lyapunov number of the map in (30) with $q = 2$ and $\kappa = \frac{1}{2}$ can be seen, respectively, in Figs. 7 and 8 ($p = 1$ and $\sigma = 0.9$).

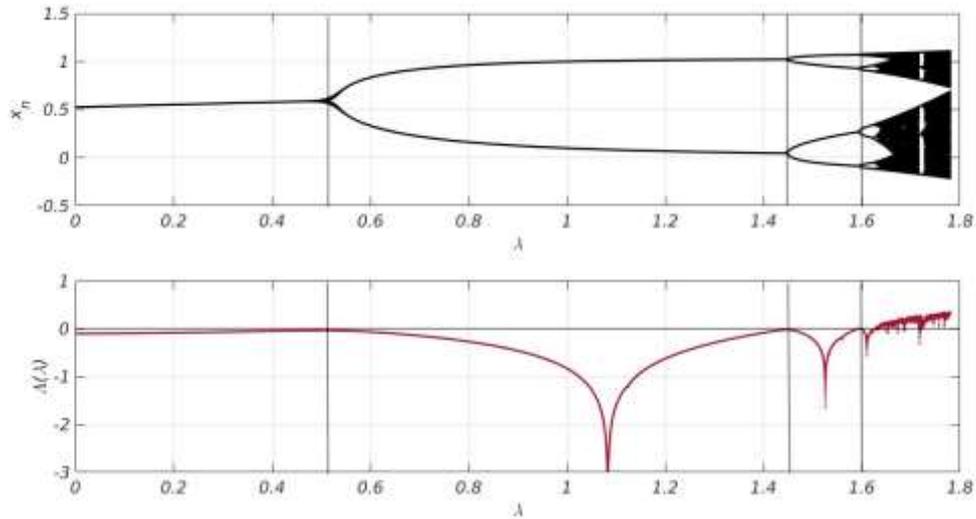

Fig. 7. Bifurcation diagram and Lyapunov number versus $\lambda$ (eq. (30) with $q = 2$, $p = 1$ and $\sigma = 0.9$).

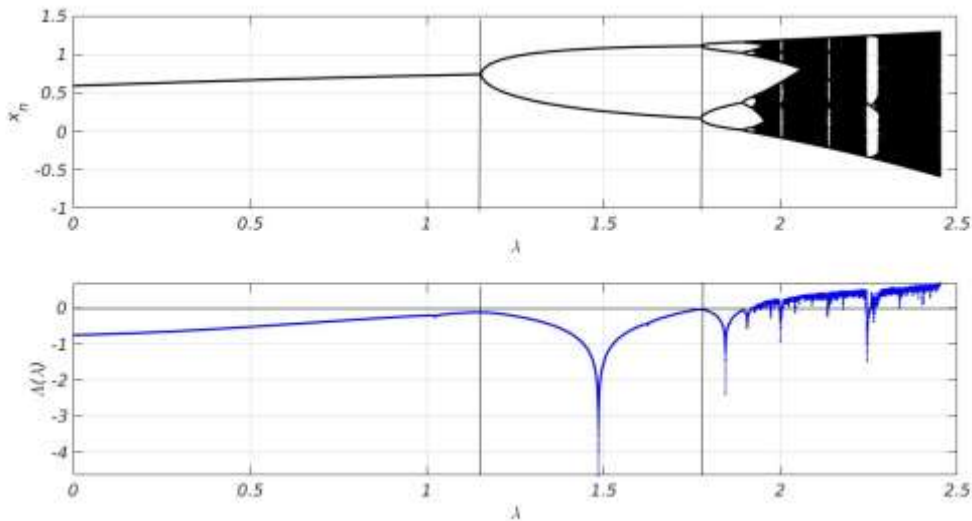

Fig. 8. Bifurcation diagram and Lyapunov number versus $\lambda$ (eq. (30) with $\kappa = 1/2$, $p = 1$ and $\sigma = 0.9$).

The chaotic regime occurs only when the Lyapunov number is positive (the Lyapunov number is zero at the bifurcation points). In [14,15] Tsallis at al showed that, at the edge of chaos (Lyapunov number positive but close to zero), the difference of two identical logistic maps with exponential decay, starting with very close initial values, grows with *q*-exponential form. It was also shown the entropy of an ensamble (the initial value is randomly chosen) of logistic maps with exponential decay also grows with *q*-exponential form. In the present section a similar analysis is done considering the logistic map with $q(\kappa)$-exponential decay given in eq. (30).

In order to analysis the sensibility to the initial conditions, we calculated $\xi(n) = \Delta x(n)/\Delta x(0)$ for the first 21 iterations, where $\Delta x(n) = x_n^1 - x_n^2$ and $\Delta x(0) = x_0^1 - x_0^2 = 10^{-12}$. The upper index 1(2) indicates map 1(2). For the map with *q*-exponential, the parameters used are $\lambda = 1.5932$, $q = 2$, $\sigma = 0.9$ and $p = 1$. Several calculations of $\xi(n)$ were done choosing randomly the initial value $x_0^1$. The *q*-log of the average value of $\xi(n)$ for different values of *q* can be seen in Fig. 9 while the $\kappa$-log of the average value of $\xi(n)$ for different values of $\kappa$ can be seen in Fig. 10.

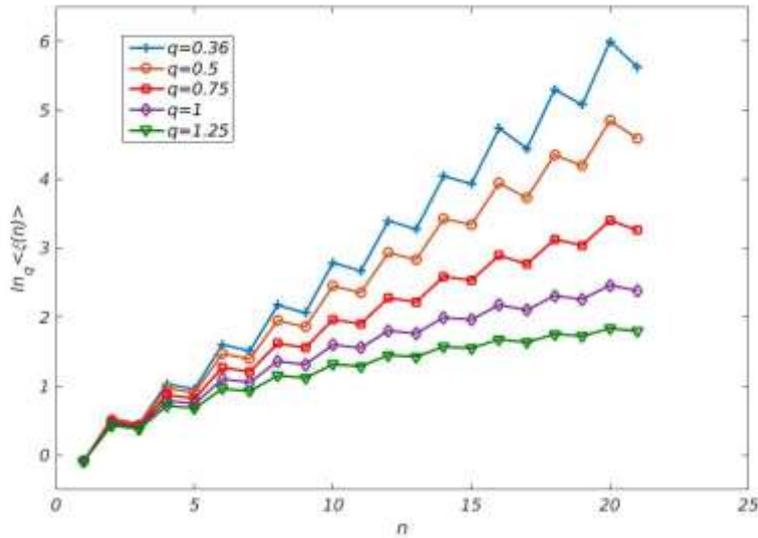

Fig. 9. $\ln_q\langle \xi(n)\rangle$ versus *n* for $q = 2$ in eq. (30).

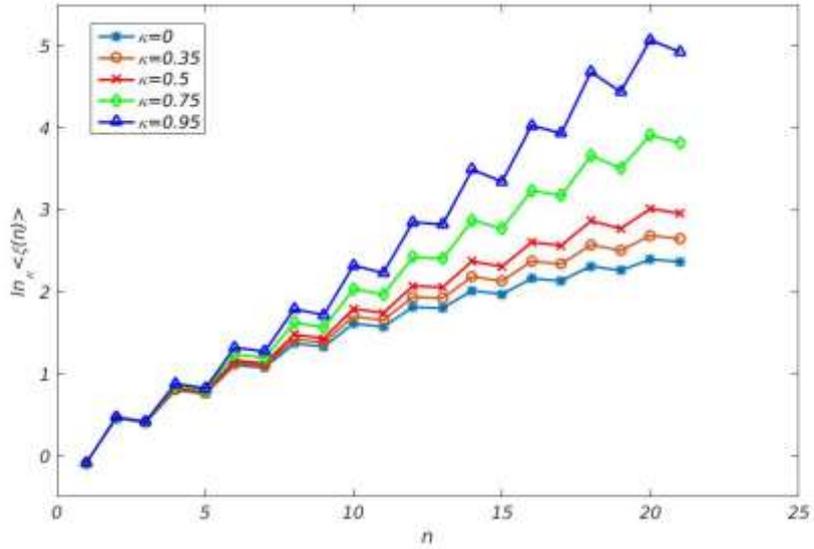

Fig. 10. $\ln_\kappa \langle \xi(n) \rangle$ versus $n$ for $q = 2$ in eq. (30).

The sawtooth behaviour seems to be interplay between the values of $q$ used in (30) and used in the $q$-log function. It does not decrease when the number of samples is increased. This is totally in contrast with the smoothness of the curves presented in [14,15] whose map uses a non-deformed exponential. Nevertheless, the more or less linear behaviour of $\ln_q \langle \xi(n) \rangle$ can still be seen for $q = 0.36$ and $\kappa = 0.95$.

The second situation considered is $q = 1.75$ in eq. (30). The curve of $\ln_q \langle \xi(n) \rangle$ versus $n$ for different values of $q$ can be seen in Fig. 11.

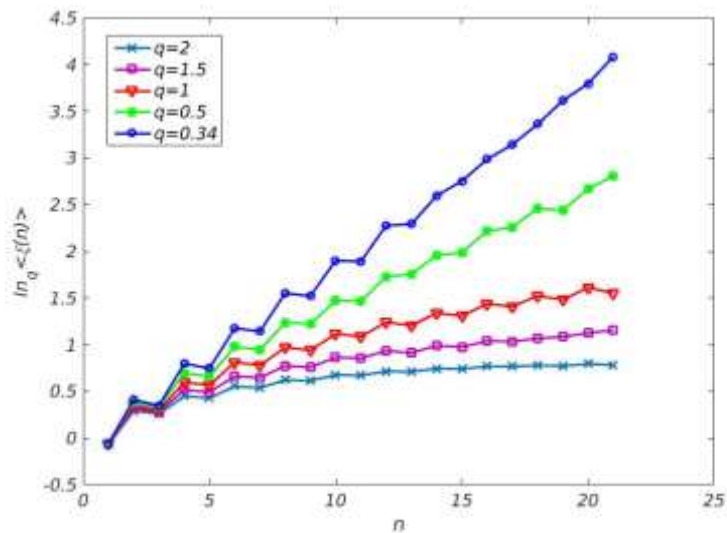

Fig. 11. $\ln_q \langle \xi(n) \rangle$ versus $n$ for $q = 1.75$ in eq. (30).

As it can be seen in Fig. 11, the 'almost' linear behaviour appears for $q = 0.34$. Table 1 shows the relation between $q$, $\lambda$ (and its respective Lyapunov number) and the best value of $q$ ($q_b$) for which $\ln_q\langle\xi(n)\rangle$ can be well fitted by a straight line.

Table 1. Relation between $q$, $\lambda$, $\Lambda(\lambda)$ and the best value of $q$ for which $\ln_q\langle\xi(n)\rangle$ can be well fitted by a straight line.

| $q$ in eq. (30) | $\Lambda$ | $\lambda$ | $q_b$ |
|---|---|---|---|
| 2 | 0.000521 | 1.5932 | 0.36 |
| 1.75 | 0.000644 | 1.773 | 0.34 |
| 1.5 | 0.000430 | 1.843 | 0.32 |
| 1.25 | 0.006611 | 1.897 | 0.31 |

Let us consider now eq. (30) with $\kappa$-exponential decay and the following parameters values: $\kappa = \frac{1}{2}$, $p = 1$, $\sigma = 0.9$ and $\lambda = 1.918$. The values of $\ln_{\kappa(q)}\langle\xi(n)\rangle$ for different values of $\kappa(q)$ can be seen in Fig. (12) (Fig. (13)).

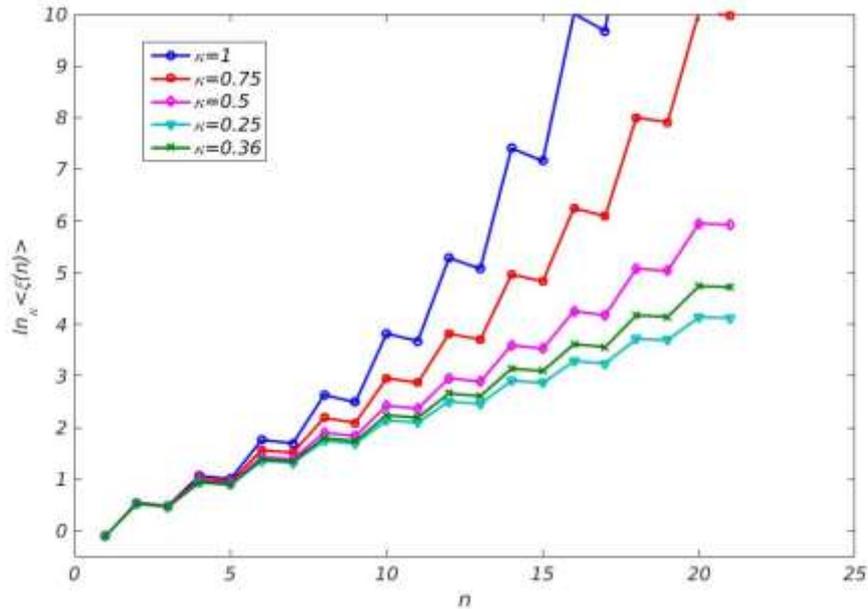

Fig. 12. $\ln_\kappa\langle\xi(n)\rangle$ versus $n$ for $\kappa = \frac{1}{2}$ in eq. (30).

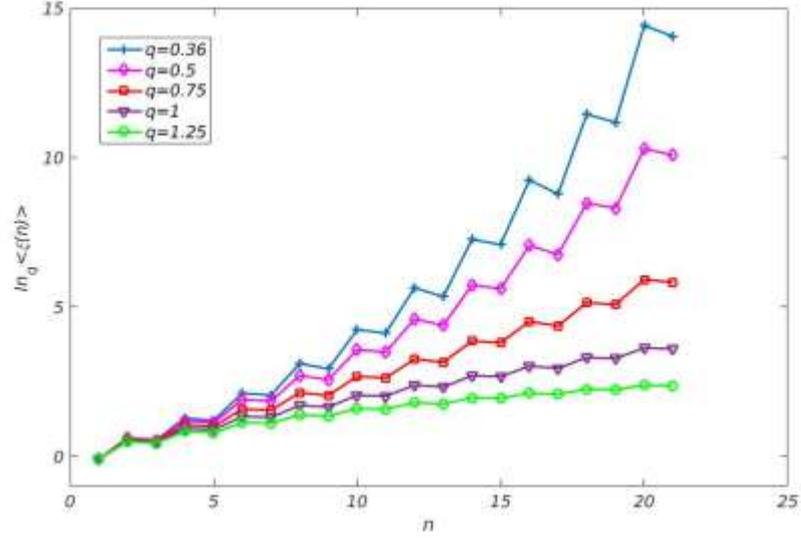

Fig. 13. $\ln_q \langle \xi(n) \rangle$ versus $n$ for $\kappa = \frac{1}{2}$ in eq. (30).

Aiming to get a linear behaviour of $q$-log and $\kappa$-log, the best values for $q$ and $\kappa$ are, respectively, $q = 0.75$ in Fig. 12 and $\kappa = 0.5$ in Fig. 13. Putting all together, it seems the logistic map with deformed exponential decay shows a 'linear-modulated' growth of $\ln_{q(\kappa)} \langle \xi(n) \rangle$ and, hence, one cannot say the growth of the distance between identical maps that start with very close initial conditions follows a $q$-exponential or $k$-exponential form.

At last, one can see in Fig. 14 the evolution of the disentropy. For $q = 2$ in eq. (30) the variable $x_n$ assumes values in the range [-0.2,1]. This interval é divided in 100 parts. Each small part of that interval is a site. The map is iterated fifty times and the whole process is repeated $K$ times. For each time the initial value is randomly chosen in the interval $[0, 2.2 \times 10^{-5}]$. The number of times the $i$-th site was visited by $x_n$ during the $m$-th iteration is stored. At the end, one has a histogram for each iteration and the disentropy of that histogram is calculated. During the first 20 iterations the dynamic of the maps is almost the same and, hence the disentropy is the same. After 20 iterations, the solutions starts to diverge and more sites are visited, decreasing the disentropy. At the end, like a random system, all sites are equally visited, making the disentropy to vanish. The curve is shown in Fig. 14. The parameter values used are $q = 2$, $p = 1$, $\sigma = 0.9$ and $\lambda = 1.6435$.

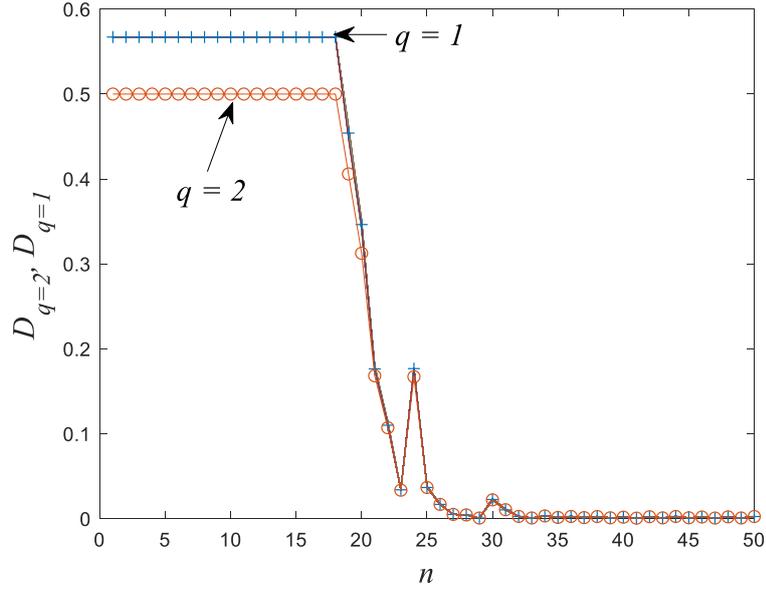

Fig. 14. Disentropy versus $n$ for $q = 2$ in eq. (30).

As one can see in Fig. 14, the decayment of the disentropy is almost linear for $q = 1$(plus signal) and $q = 2$ (ball). The growth of the disentropy at iteration number 24 is not smoothed with the increase of samples but it can be eliminated if the number of sites increase to 10000, for example. However, in this case the maximal value of the disentropy decreases.

## 3. Ordinary differential equation with $q$-exponential decay

The set of differential equations that model the motion of a projectile in a resistive medium (the resistance is proportional to the velocity via constant $k$) are

$$\frac{d^2 y}{dt^2} + k\frac{dy}{dt} + g = 0 \tag{34}$$

$$\frac{d^2 x}{dt^2} + k\frac{dx}{dt} = 0. \tag{35}$$

The solutions of eqs. (34) and (35) are, respectively [16],

$$y(t) = -\frac{tg}{k} - \alpha + \alpha e^{-kt} \qquad (36)$$

$$x(t) = \frac{1}{k}\left[v\cos(\theta)\left(1 - e^{-kt}\right)\right] \qquad (37)$$

$$\alpha = \frac{g}{k^2}\left(-1 - \frac{kv\sin(\theta)}{g}\right). \qquad (38)$$

In eqs. (36)-(37) $v$ is the velocity at time $t = 0$ and $\theta$ is the elevation angle. The range is the value of $x(t = t_{end})$ where $t_{end}$ is obtained from $y(t = t_{end}) = 0$. Hence, using (36) one gets ($k^2\alpha/g > 1$)[16]

$$t_{end} = -\frac{k\alpha}{g} + \frac{1}{k}W\left(\frac{k^2\alpha}{g}\exp\left(\frac{k^2\alpha}{g}\right)\right), \qquad (39)$$

and the range is obtained using eq. (39) in eq. (37).

Now, let us assume the vertical motion is governed by the following non-linear differential equation

$$\frac{\alpha e_q^{-kt}}{q}\frac{d^2y}{dt^2} - \left(\frac{dy}{dt}\right)^2 - 2\frac{g}{k}\frac{dy}{dt} - \frac{g^2}{k^2} = 0, \qquad (40)$$

whose solution is

$$y(t) = -t\frac{g}{k} - \alpha + \alpha e_q^{-kt}. \qquad (41)$$

In this case, $t_{end}$ is given by the stable point of the map

$$t_{n+1} = -\frac{1}{k}W_q\left(-t_n^2\frac{g}{\alpha} - kt_n\right). \qquad (42)$$

Figure 15 shows (in a unitless example) the plot of eq. (41) ($k = 0.1$, $g = 32.2$, $v = 50$ and $\theta = \pi/4$) for three different values of $q$ (0.5, 1, 2). The values of $t_{end}$ obtained using eq. (42) are: $q = 0.5 \to 3.957$, $q = 1 \to 2.121$, $q = 2 \to 1.097$.

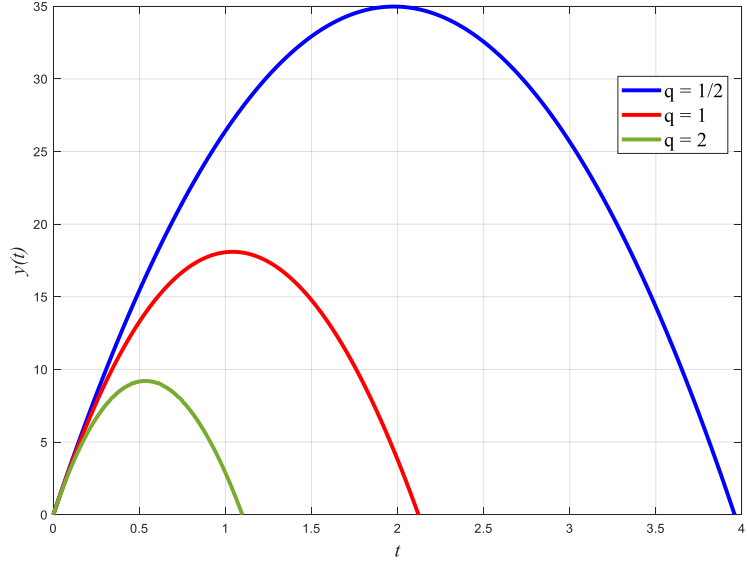

Fig. 15. $y(t)$ versus $t$ for three different values of $q$ (0.5, 1, 2).

Using $q = 2$ (eq. (14.a)) and $t_{n+1} = t_n = t_{end}$ in (42), one finds easily that $t_{end} = -(k\alpha/g+1/k)$ = 1.097.

## 4. Conclusions

The Lambert-Tsallis $W_q(z)$ and Lambert-Kaniadakis $W_\kappa(z)$ functions are important mathematical tools that can be used in the study of problems modelled by difference and differential equations with deformed exponential decay. In particular we showed the calculation of the stable point and the evolution of the disentropy in the weak chaotic regime of a logistic map with deformed exponential decay. The strong decayment of disentropy is clearly a signature of a chaotic regime. We also introduced a new differential equation whose solution is similar to the solution of the motion of a projectile with a given initial velocity and elevation angle in a resistive medium. However, our solution employs the Tsallis exponential instead the non-deformed exponential. In this case, the Lambert-Tsallis function was used to calculate the range of the projectile.

# References

1. R. M. Corless, G. H. Gonnet, D. E. G. Hare, D. J. Jeffrey and D. E. Knuth, *On the Lambert W function*, Advances in Computational Mathematics, vol. 5, 329 – 359 (1996).
2. S. R. Valluri, D. J. Jeffrey, R. M. Corless, Some applications of the Lambert W function to Physics, Canadian Journal of Physics, vol. 78 n° 9, 823-831 (2000).
3. D. C. Jenn, Applications of the Lambert W function in Electromagnetics, IEEE Antennas and Propagation Magazine, vol. 44, n° 3 (2002).
4. F. C.-Blondeau and A. Monir, Numerical evaluation of the Lambert *W* function and application to generation of generalized Gaussian noise with exponent ½, IEEE Transactions on Signal Processing, vol. 50, no. 9, 2160-2165 (2002).
5. D. Veberic, Having fun with Lambert $W(x)$ function, GAP-2009-114 [Online]. Available: http://arxiv.org/abs/1003.1628.
6. K. Roberts, S. R. Valluri, Tutorial: The quantum finite square well and the Lambert W function, Canadian Journal of Physics, vol. 95, no. 2, 105-110 (2017).
7. G. B. da Silva and R.V. Ramos, The Lambert-Tsallis $W_q$ function, Physica A, **525**, 164-170 (2019). https://doi.org/10.1016/.physa.2019.03.046.
8. C. Tsallis, Possible generalization of Boltzmann-Gibbs statistics, J. Stat. Phys. 52, 479 (1988).
9. G. Kaniadakis, Statistical mechanics in the context of special relativity. Physical Review E, v. 66, n. 5, p. 056125 (2002).
10. E. MF Curado, and C. Tsallis. "Generalized statistical mechanics: connection with thermodynamics." *Journal of Physics A: Mathematical and General* 24.2 (1991): L69.
11. R. R. Serrezuela, O. F. Villar, J. R. Zarta, Y. H. Cuenca, The k-Exponential Matrix to solve systems of differential equations deformed, Global Journal of Pure and Applied Mathematics, Vol. 12, N° 3, 1921-1945 (2016).
12. J. L. E. da Silva, G. B. da Silva. R. V. Ramos, The Lamber-Kaniadakis $W_\kappa$ function, Phys. Lett. A, (2019). DOI: 10.1016/j.physleta.2019.126175
13. Braverman, E., and S. H. Saker. "On a difference equation with exponentially decreasing nonlinearity", *Discrete Dynamics in Nature and Society*, vol. 2011, 1-17, (2011). DOI: 10.1155/2011/147926.

14. V. Latora, M. Baranger, A. Rapisarda, C. Tsallis, The rate of entropy increase at the edge of chaos, Physics Letters A, 273, 97–103 (2000).

15. G. F. J. Ananos, C. Tsallis, Ensemble averages and nonextensivity at the edge of chaos of one-dimensional maps. Physical review letters, 93(2), 020601 (2004).

16. E. W. Packel and D. S. Yuen, Projectile motion with resistance and the Lambert W Function, The College Mathematics Journal, vol. 35, No. 5, 337-350 (2004).